\documentclass[twocolumn,showpacs,preprintnumbers,prl,floatfix]{revtex4}
\usepackage{amssymb}
\usepackage{graphicx}
\usepackage{dcolumn}
\usepackage{bm}

\begin{document}

\title{Backflow Effect on Spin Diffusion 
Near
Ferromagnet-Superconductor Interface}
\author{M. Faiz, R.P. Panguluri}
\affiliation{Department of Physics and Astronomy, Wayne State University, Detroit,
Michigan 48201}
\author{B. Balke}
\affiliation{Institute for Materials Science, University of Stuttgart, 70569 Stuttgart, Germany}
\author{S. Wurmehl}
\affiliation{Institute for Materials Research IFW, 01069 Dresden, Germany}
\author{C. Felser} 
\affiliation{Max Planck Institute for Chemical Physics of Solids, 01187 Dresden, Germany }
\author{A. G. Petukhov}
\affiliation{Google Inc., Venice, California 90291, USA}
\author{B. Nadgorny}
\affiliation{Department of Physics and Astronomy, Wayne State University, Detroit,
Michigan 48201}
\date{\today }

\begin{abstract}
The behavior of spin propagation in metals in various measurement schemes is
shown to be qualitatively different than a simple exponential decay - due to
the backflow effect on spin diffusion in the presence of interfaces. To
probe this effect we utilize the spin sensitivity of an Andreev contact
between gold films of variable thickness deposited on top of a spin
injector, Co$_{2}$Mn$_{0.5}$Fe$_{0.5}$Si, with the spin polarization of
approximately 45\%, and Nb superconducting tip. While the results are
consistent with gradually decaying spin polarization as the film thickness
increases, the spin diffusion length in Au found to be 285 nm, is more than
two times larger that one would have obtained without taking the backflow
effect into account.  
\end{abstract}

\author{}
\maketitle

Processes of spin injection and spin accumulation are of fundamental
importance for operation and underlying physics of spintronic devices~\cite%
{ZuticREVMOD}. After it was realized that a spin polarized current can
induce 
non-equilibrium spin populations of both nuclear \cite{Overhauser} and
electronic \cite{Feher} subsystems in a 
normal (non-magnetic) metal, the related problem of spin injection from a
ferromagnet (F) into a normal metal (N) was considered by Aronov \cite%
{Aronov}. 
Johnson and Silsbee~\cite{JohnsonS} performed 
the first measurements of spin relaxation in a purely electronic subsystem.
These experiments utilized the so-called lateral non-local geometry to
determine a spin diffusion length 
in aluminum 
by probing a difference 
between chemical potentials of the two spin subbands. A more convenient
version of this technique was later 
adopted for F/N/F structures \cite{Johnson}, and has been 
further developed by Jedema \emph{et al.}~\cite{vanWees}. Another means to
determine 
spin-diffusion length in metals is to analyze the thickness dependence of
the current-perpendicular-to plane (CPP) giant magnetoresistance (GMR)
effect \cite{BassPratt,BP}. 
Finally, an optical technique based on measuring the spin accumulation via
the Kerr effect has 
been successfully implemented by Crooker~\emph{et al.}~\cite{Crowell}. \qquad

Most of the measurement techniques described above use the implicit
assumption that spin in a normal metal decays exponentially with distance.
While in the case of spin injection into a normal metal of infinite
thickness this assumption is correct, the presence of a spin selective
interface within a distance that is comparable to the spin diffusion length
would modify this dependence in any real measurements. Indeed, a spin
selective interface imposes different boundary conditions for spin-up and
spin-down electrons, thus resulting in a backflow of spin polarized
electrons away from that interface.

The backflow effect exists in the case of an N/F interface and thus have
significant implications for the description of spin accumulation and spin
propagation in GMR\ devices, but it arguably can be the most pronounced in
the case of N/S interface. At the energies below the superconducting gap $%
\Delta $ and temperatures far enough from the superconducting transition
temperature $T_{c}$, 
Andreev reflection \cite{Andreev} is the dominant process \cite{lowbarrier}
that allows quasiparticle current propagation from a normal metal into a
superconductor by converting quasiparticles with opposite spins into Cooper
pairs. Any asymmetry in the quasiparticle spin balance, that may exist, for
example in a ferromagnet, would reduce the probability of such a process and
consequently the conductance across the interface~\cite{DeJong}. Based on
this property of Andreev reflection at an F/S interface it has been shown
that the junction conductance is sensitive to the values of spin
polarization in a ferromagnet \cite{US,Buhrman}. Similarly, a spin current
injected into a normal metal should be sensitive to the same Andreev
reflection mechanism due the non-equilibrium spin accumulation near an N/S
interface. Such spin accumulation will gradually decrease as we increase the
thickness of the N-layer~\cite{Klapwijk}.

\begin{figure}[tbh]
\includegraphics[width=\linewidth]{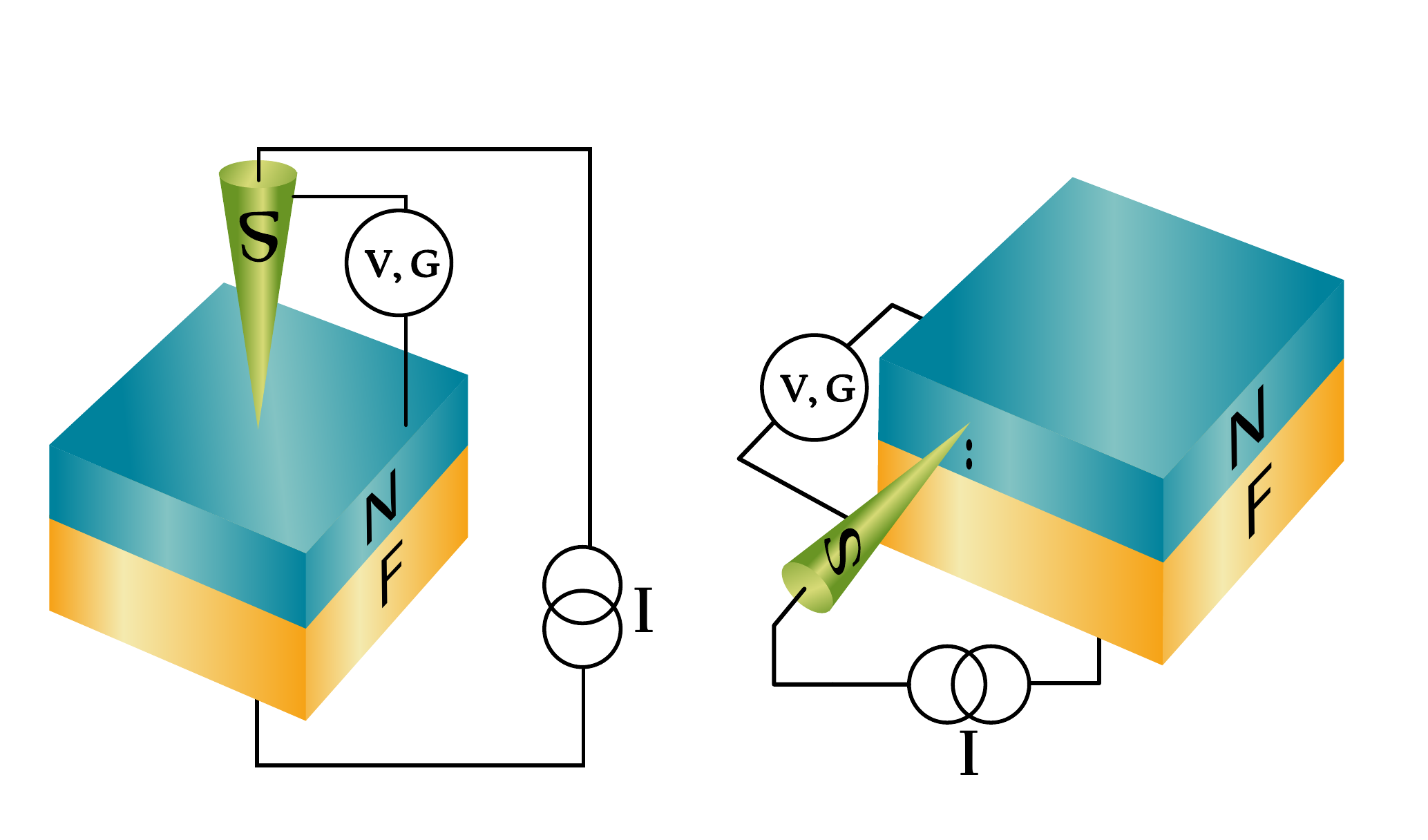}
\caption{(Color online.) Schematics of the 
PCAR experiment presented in this work (left) and the next generation PCAR experiment (right)}
\label{schematics}
\end{figure}

In this Letter we propose to use Point Contact Andreev Reflection (PCAR)
spectroscopy to investigate the backflow effect on spin diffusion and spin
accumulation by exploiting the dependence of the magnitude of this effect on
metal thickness, \ as shown in Fig.~\ref{schematics}. In particular, we use
spin injection from a highly spin polarized Heusler alloy, Co$_{2}$Mn$_{0.5}$%
Fe$_{0.5}$Si into gold films of different thicknesses to observe a gradual
decay of spin polarization in Au. We formulate a phenomenological
description of such transport in 
a diffusive regime to determine the spin diffusion length $L_{N}$ in gold
and demonstrate that a combination of the PCAR\ technique with the proper
phenomenological theory could result in an alternative electrical technique
for probing spin diffusion length in normal metals.

As most of the Heusler alloys \cite{deGroot} Co$_{2}$Mn$_{0.5}$Fe$_{0.5}$Si
has a high ($\sim $ 1000K) Curie temperature and is believed to be fairly
highly spin polarized
. 
The samples of the Heusler alloy Co$_{2}$Mn$_{0.5}$Fe$_{0.5}$Si were
fabricated by arc melting from stoichiometric ratio of constituents in an
argon atmosphere of 10$^{-4}$ mbar. After subsequent annealing of the
polycrystalline ingots in an evacuated quartz tube at 1273K for 21 days the
samples with the Heusler type L21 structure were obtained, as was verified
by X-ray powder diffraction (XRD) using Mo K$_{\alpha }$ excitation. Flat
disks were then cut from the ingots and polished before removing the native
oxide by Ar$^{+}$ ion bombardment. The sample composition was further
verified by X-ray photoemission (ESCA) with no impurities detected. 
Gold films of 99.99\% purity and variable thicknesses (from 7 nm to 475nm)
were then deposited on the polished surface of the disks by thermal
evaporation in vacuum, immediately followed by the PCAR measurements.

\begin{figure}[tbph]
\includegraphics[width=\linewidth]{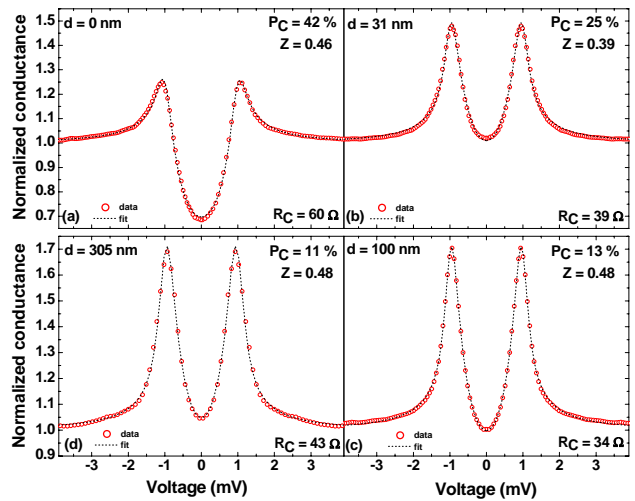}
\caption{I-V curves for Andreev reflection experiments}
\label{conductance}
\end{figure}

The measurements of the structure shown in Fig.~\ref{schematics} were
performed in the point contact geometry with Nb superconducting tips. The
tips were fabricated by the standard electrochemical etching of 250 ${%
\mu
\text{m}}$ Nb wire, as described in Ref.~\cite{FaizAPL}. \ Using freshly
etched Nb tips and oxide-free Au film helped to facilitate the establishment
of a stable contact (on the order of 50-100 $\Omega $), typically without
the need of further adjustments, thus largely alleviating any concerns of
tip-film mechanical interference; additionally post-measurement microscopy
of the contact area was performed. The current--voltage ($I-V$) and the
differential conductance $dI/dV$ measurements were performed by a standard
four-probe technique as described in detail in Ref.~\cite{MnAs} in the
temperature range of 1.2-- 4.2 K. 
The $dI/dV$ curves are analyzed with the appropriately modified \cite{MGN}
Blonder-Tinkham-Klapwijk (BTK) weak coupling theory \cite{BTK}, with two
fitting parameters, the value of spin polarization, $P$ and the interface
scattering strength $Z$. \ First, we determined the spin polarization for
bare Co$_{2}$Mn$_{0.5}$Fe$_{0.5}$ as an average over 15 different junctions; 
$P$ was found to be approximately 44 $\pm $ $3$\%, somewhat lower than for Co%
$_{2}$FeSi alloy described in earlier work \cite{FaizAPL}. For gold films
deposited onto Co$_{2}$Mn$_{0.5}$Fe$_{0.5}$ at least ten different junctions
were analyzed for each film thickness. In most cases either no or a weak $%
P(Z)$ dependence was observed, in the latter case $P$ was extrapolated the
low $Z$ limit. In Fig.~\ref{conductance} four characteristic conductance
curves for progressively thicker Au films are shown; the results are
consistent with the notion of spin polarized current gradually decaying as
the Au film thickness increases.

Most of the experiments on the spin injection into metals or semiconductors
rely on a diffusive description of the spin transport. This is based on the
fact that the spin diffusion length $L_N$ in a particular sample is related
to the value of the elastic mean free path $l$ as $L_N=l\sqrt{\tau_s/\tau_p}$%
, where $\tau_s$ and $\tau_p$ are the spin and momentum relaxation times
respectively. It is generally assumed that $\tau_s\gg\tau_p$, which, in
turn, justifies a description of the spin relaxation process within the
diffusive transport limit. Indeed, in most metals the spin diffusion length
was found to be roughly on the order of several hundred nanometers at low
temperatures~\cite{BP}, which is definitely larger than the typical values
of the elastic mean free path.

As no spin current can propagate below the gap inside the superconductor due
to the fact that only Cooper pairs with $S=0$ can be present there \cite%
{Klapwijk}, we will assume that the spin current goes to zero at the N/S
interface, neglecting any possible proximity effects. In addition, we will
use a 1D model to describe the spin current through the system. The validity
of these assumption and their possible effect on our results will be
discussed later. Our main conjecture is that the spin polarization $P(w)$
measured in the Andreev reflection experiments is proportional to the
splitting of the electrochemical potentials at the normal metal -
superconductor (N-S) interface $\Delta \zeta _{N}(w)$.

\begin{figure}[tbph]
\includegraphics[width=0.95\linewidth]{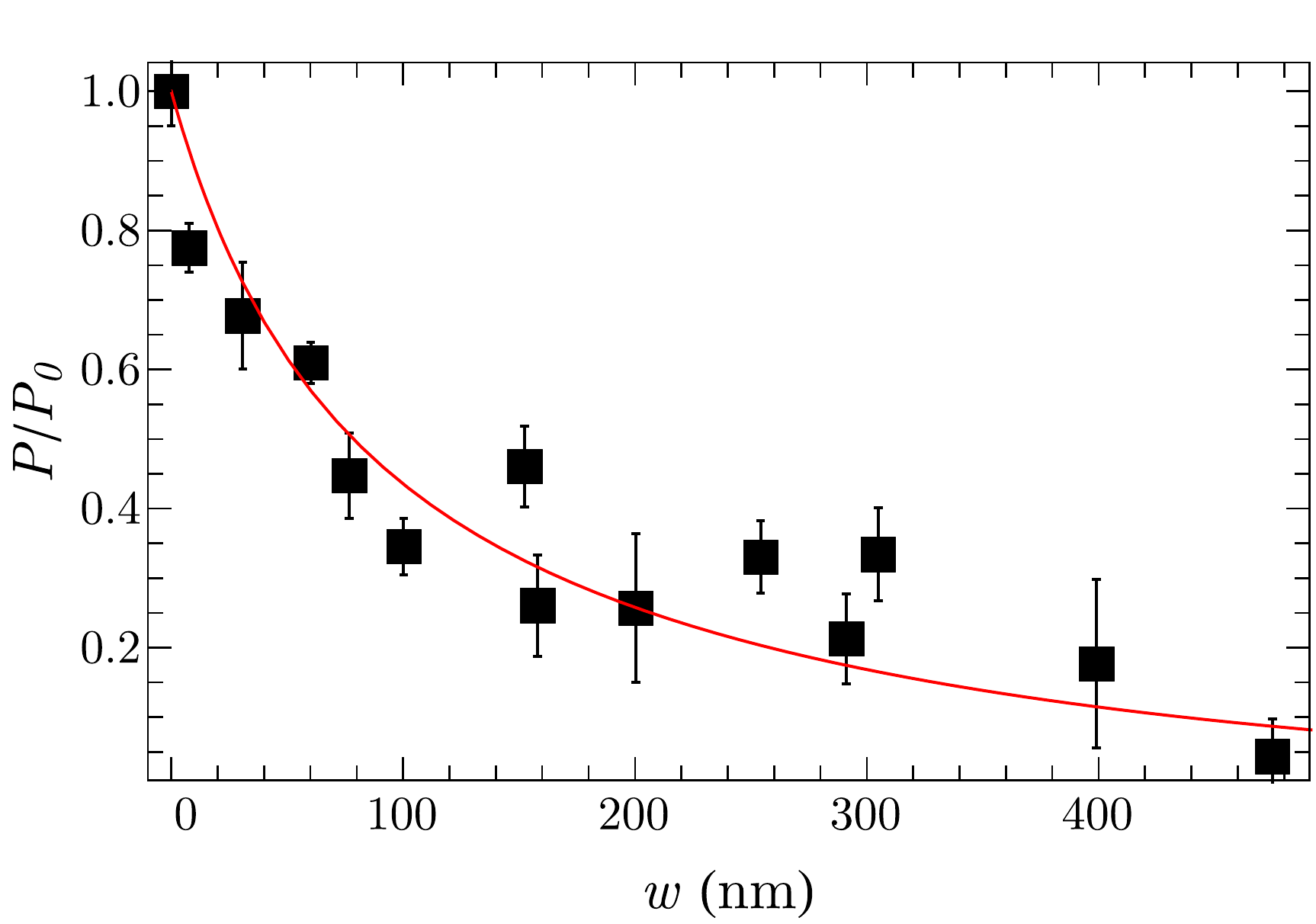}
\caption{Thickness dependence of $P/P_0$. 
Squares -- experimental data; red
solid line -- best fit using Eq.~(\ref{P}) 
with $\kappa$=3.5
and $L_N$=286~nm.}
\label{BestFit}
\end{figure}

The splitting $\Delta \zeta _{N}(x)$ is a solution of a diffusion equation: $%
\Delta \zeta _{N}(x)=A_{w}\exp (-x/L_{N})+B_{w}\exp (x/L_{N}),$where $L_{N}$
is the spin diffusion length of a normal metal and the coefficients $A_{w}$, 
$B_{w}$ must be determined from the boundary conditions. The spin
polarization of the current density can be expressed through $\Delta \zeta
_{N}(x)$ as: 
\begin{equation}
\gamma (x)=\frac{j_{\uparrow }-j_{\downarrow }}{j}=\frac{\sigma _{N}}{2j}%
\frac{d\Delta \zeta _{N}(x)}{dx},  \label{gamma}
\end{equation}%
where $\sigma _{N}$ is the bulk conductivity of the normal metal. Using the
boundary condition at $N/S$ interface $\gamma (w)=0$ we obtain $%
B_{w}=A_{w}\exp (-2w/L_{N})$ and 
\begin{equation}
\Delta \zeta _{N}(w)=\frac{\Delta \zeta _{N}(0)}{\cosh (w/L_{N})},
\label{zetaW}
\end{equation}%
where $\Delta \zeta (0)$ is the splitting of the electrochemical potentials
at $F/N$ interface. We note that $\Delta \zeta _{N}(0)$ depends on $w$ due
to the positive feedback exponent. To find $\Delta \zeta _{N}(0)$ we will
use Rashba's boundary condition \cite{Rashba:2000vy}: 
\begin{equation}
\Delta \zeta _{N}(0)-\Delta \zeta _{F}(0)=2jr_{c}\left[ \gamma (0)-\gamma
_{c}\right]  \label{boundary1}
\end{equation}%
Here $\Delta \zeta _{F}(x)$ is the splitting of the electrochemical
potentials in the ferromagnet, $r_{c}=(\Sigma _{\uparrow }+\Sigma
_{\downarrow })/(4\Sigma _{\uparrow }\Sigma _{\downarrow })$, $\gamma
_{c}=(\Sigma _{\uparrow }-\Sigma _{\downarrow })/(\Sigma _{\uparrow }+\Sigma
_{\downarrow })$, and $\Sigma _{\uparrow ,\downarrow }$ are the contact
conductances. Another boundary condition is the continuity of the spin
current across $F/N$ interface~\cite{Rashba:2000vy}: 
\begin{equation}
\sigma _{N}\Delta \zeta _{N}^{\prime }(0)-4(\sigma _{\uparrow }\sigma
_{\downarrow }/\sigma _{F})\Delta \zeta _{F}^{\prime }(0)=2\gamma _{F}j,
\label{boundary2}
\end{equation}%
where $\gamma _{F}=(\sigma _{\uparrow }-\sigma _{\downarrow })/\sigma _{F}$, 
$\sigma _{F}=\sigma _{\uparrow }+\sigma _{\downarrow }$, and $\sigma
_{\uparrow ,\downarrow }$ are the bulk conductivities of the ferromagnet. We
note that in the semi-infinite ferromagnet $\Delta \zeta _{F}(x)=C\exp
(x/L_{F})$, where $L_{F}$ is the ferromagnet spin diffusion length. This
implies that $\Delta \zeta _{F}^{\prime }(0)=\Delta \zeta _{F}(0)/L_{F}$.
Also $\Delta \zeta _{N}^{\prime }(0)=-\tanh (w/L_{N})\Delta \zeta
_{N}(0)/L_{N}$. Substituting these formulas in Eqs~(\ref{boundary1}) and (%
\ref{boundary2}), eliminating $\Delta \zeta _{F}(0)$, and using Eq.~(\ref%
{gamma}) we finally obtain: 
\begin{equation}
\gamma (0)=\frac{\gamma _{c}r_{c}+\gamma _{F}r_{F}}{r_{F}+r_{c}+r_{N}/\tanh
(w/L_{N})}
\end{equation}

and 
\begin{equation}
\Delta \zeta _{N}(0) = \frac{2|j|(\gamma _{c}r_{c}+\gamma _Fr_F)r_N}{%
(r_{c}+r_{F})\tanh (w/L_N)+r_N}  \label{zeta0}
\end{equation}%
Here we introduced the resistances $r_{F}=L_{F}\sigma _{F}/(4\sigma
_{\uparrow }\sigma _{\downarrow })$, and $r_{N}=L_{N}/\sigma _{N}$. Using
Eqs.~(\ref{zetaW}) and (\ref{zeta0}) we can calculate the spin polarization
at the N/S interface, 
$P=P(w)\propto \Delta \zeta_N (w)$, which yields: 
\begin{equation}  \label{P}
P(w)=\frac{P_0}{\kappa\sinh(w/L_{N})+\cosh(w/L_{N})},
\end{equation}%
where $\kappa=(r_{c}+r_{F})/r_{N}$ and $P_0\propto \gamma_c
(r_c/r_N)+\gamma_F (r_F/r_N),$ 
is the limiting value of the spin polarization at small $w$. 
The results of our fitting procedure are shown in Fig~\ref{BestFit}, with $%
L_N\simeq$~285~nm and $\kappa\simeq$~3.5.

\begin{figure}[tbph]
\includegraphics[width=\linewidth]{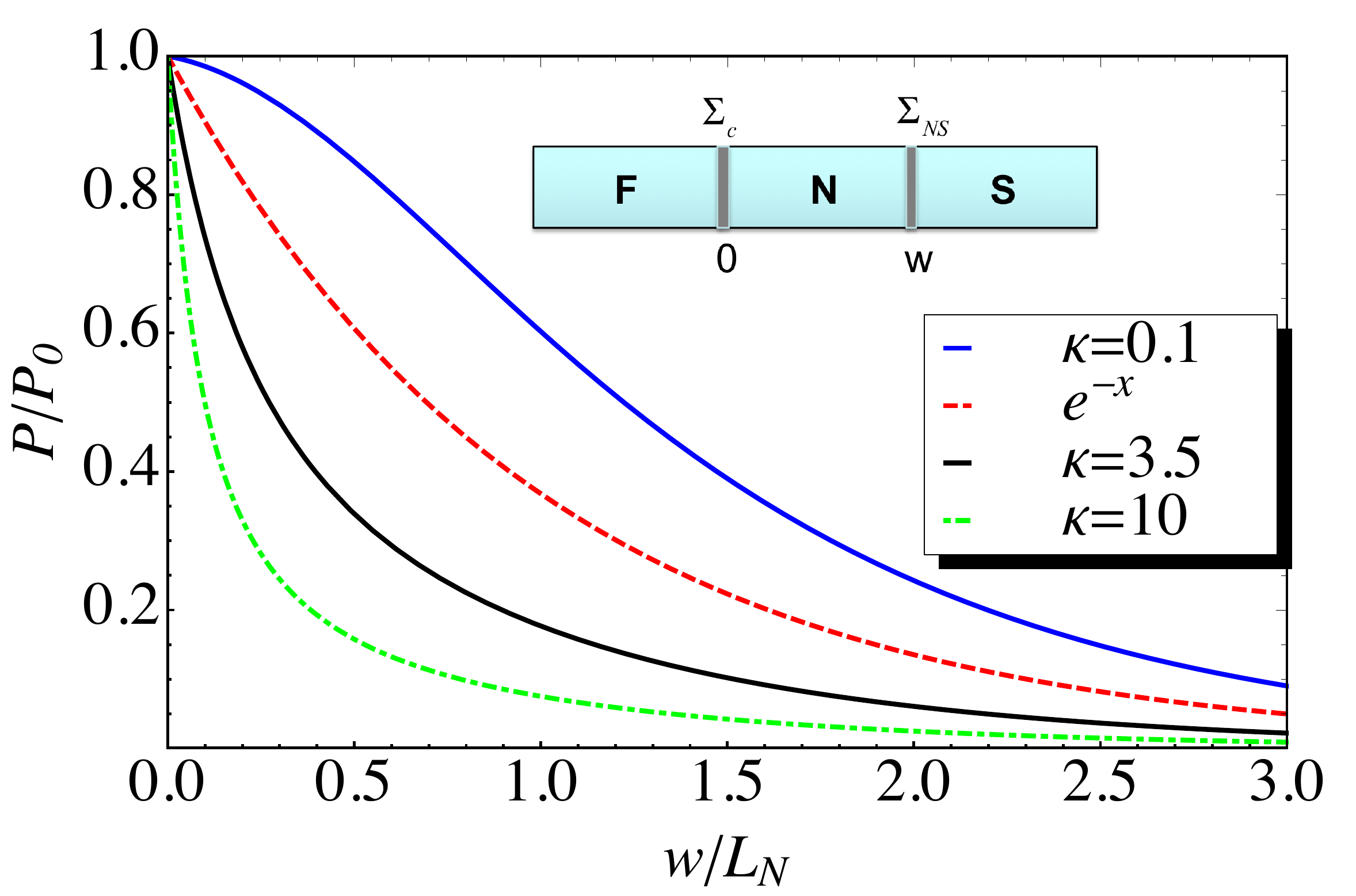}
\caption{Normalized spin polarization $P/P_0$ for different values of $\kappa$.}
\label{P_fig}
\end{figure}

The qualitative dependence of $P/P_0$ for three different values of $\kappa$
is shown in Fig.~\ref{P_fig}. As can be seen from the plot, the thickness
dependence of $P$ is much sharper than the simple exponential dependence, $%
P\propto\exp(-w/L_N)$, which is often used to fit the spin diffusion data.
Indeed, at small $w$ , $P\simeq 1-\kappa w/L_N$ rather than $1-w/L_N$. It
means that for $\kappa >1$ the spin polarization in Eq.~(\ref{P}) decays
faster than the simple exponent. Thus, if we attempted to fit our data with
a simple exponential dependence we would obtain $L_{\text{eff}}\simeq
L_N/\kappa$. In our case, this is about three times smaller than the actual
value. The best fit with the simple exponential dependence gives $L_{\text{%
eff}}\sim$~130 nm (see Fig.~\ref{BestFit}). In addition, a na\"{\i}ve
interpretation would give different values of the apparent spin diffusion
length for different ferromagnetic spin injectors and F/N interfaces of
different quality, which is obviously a non-physical result.


Eq.~(\ref{P}) is valid at low temperatures when Andreev reflection dominates
the transport across the interface. At higher temperatures we have to take
into account the thermally activated tunneling of quasiparticles, which
leads to a non-zero spin current at the N/S interface. Following Takahashi 
\emph{et al.}~\cite{Takahashi} let us introduce the (spin-independent in our
case) tunnel conductance for the N/S interface $\Sigma_{NS}=\Sigma_N\chi(T)$
where $\Sigma_N$ is the tunnel conductance between the two normal metals
(i.e. above the superconductivity threshold $T_c$) and $\chi(T)$ is the
so-called Yosida function~\cite{Takahashi} describing increase of the
tunneling conductance as the temperature rises from 0 to $T_c$. 
\begin{equation}  \label{Yosida}
\chi(T)=2\int_\Delta^\infty\frac{E_{\bm k}}{\sqrt{E_{\bm k}^2-\Delta^2}}%
\left(-\frac{\partial f_0} {\partial{E_{\bm k}}}\right)dE_{\bm k},
\end{equation}
where $f_0(E_{\bm k})$ is the Fermi distribution function and $E_{\bm k}=%
\sqrt{\xi_{\bm k}^2+\Delta^2}$ is the quasi-particle energy with $\xi_{\bm %
k} $ being a one-electron energy relative to the chemical potential of the
superconductor.

\begin{figure}[tbph]
\includegraphics[width=1.05\linewidth]{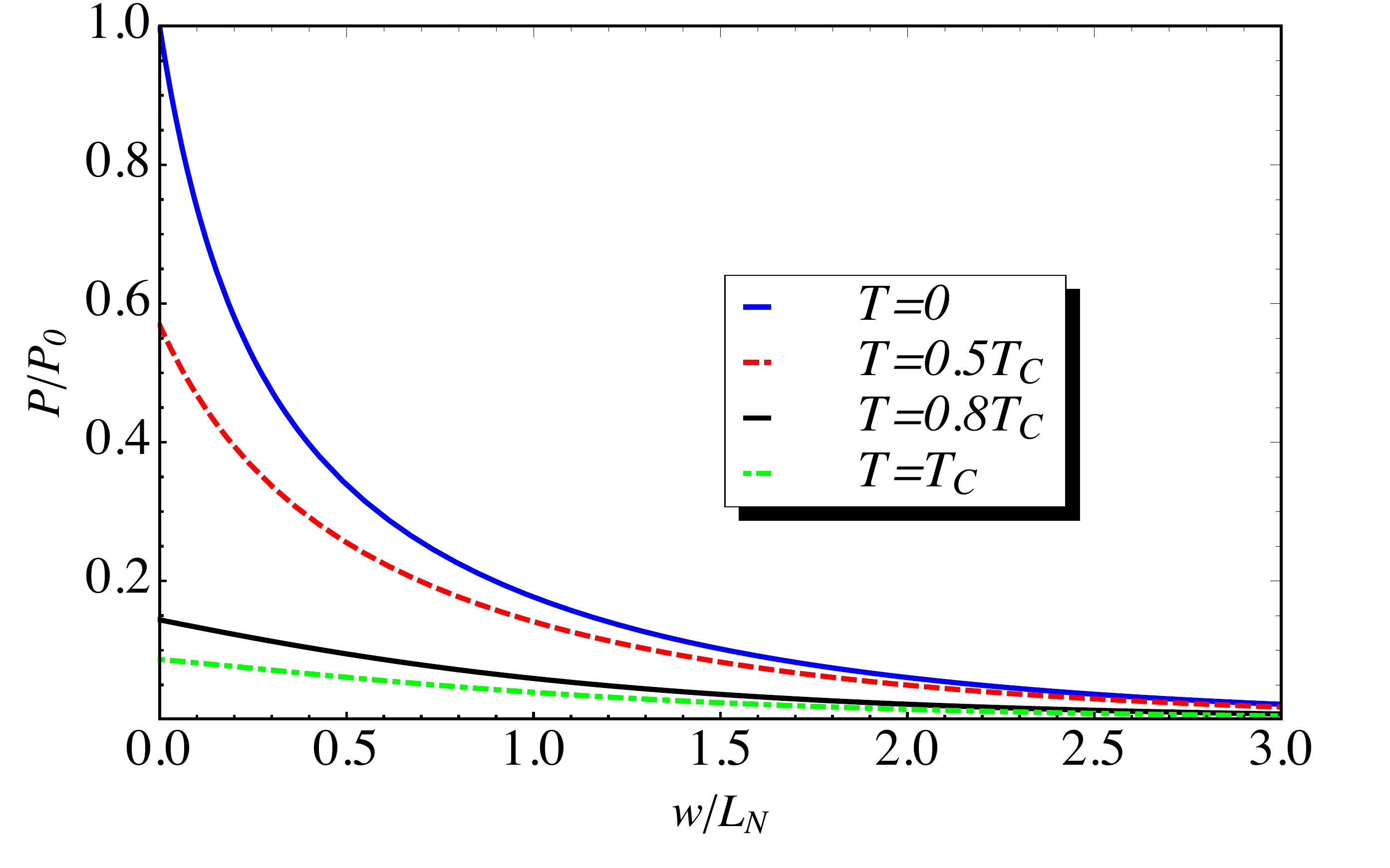}
\caption{Spin polarization $P(w,T)/P_{0}$ (Eq.~(\protect\ref{P-temp})) for $%
\protect\kappa =3.5$ and $\protect\mu =3$.}
\label{P-T}
\end{figure}

\begin{figure}[tbph]
\includegraphics[width=\linewidth]{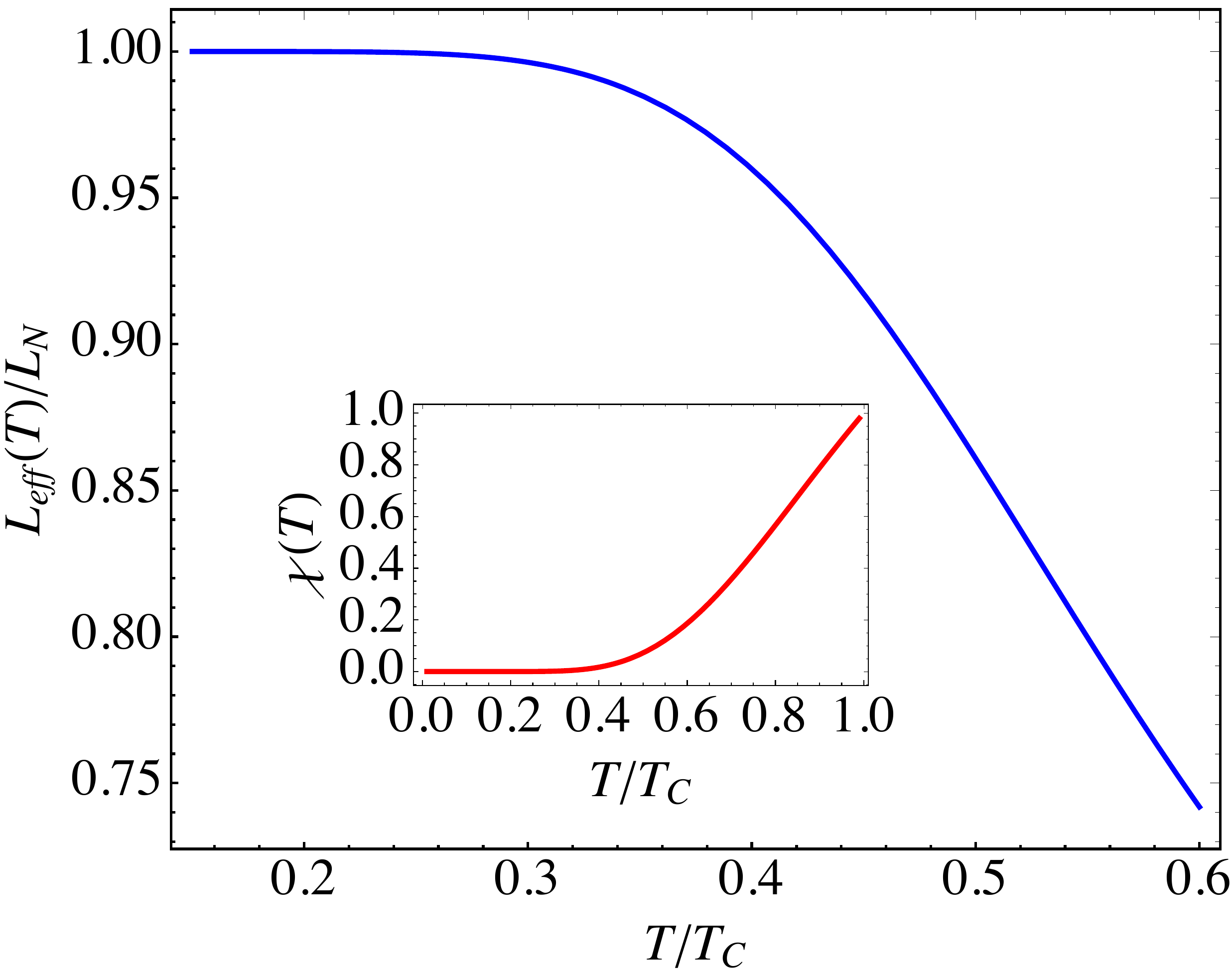}
\caption{Apparent spin diffusion length $L_{eff}(T)$ for $\protect\kappa=3.5$
and $\protect\mu=3$. Inset: Yosida function (Eq.~(\ref{Yosida}))}
\label{L-T}
\end{figure}

In the absence of the spin-flip transition at the N/S interface and in
S-region the boundary condition $\gamma (w)=0$ has to be replaced with~\cite%
{Rashba:2000vy,Takahashi}: 
\begin{equation}
2j\gamma (w)=-\Sigma _{N}\chi (T)\Delta \zeta (w)  \label{Yosida-bc}
\end{equation}%
Using the boundary condition (\ref{Yosida-bc}) we can repeat the above
calculations and obtain: 
\begin{equation}
P(w,T)=\frac{P_{0}\left[ 1+\kappa \mu \chi (T)\right] ^{-1}}{g(T)\sinh
(w/L_{N})+\cosh (w/L_{N})},  \label{P-temp}
\end{equation}%
where $\mu =r_{N}\Sigma _{N}$ and 
\begin{equation}
g(T)=\frac{\kappa +\mu \chi (T)}{1+\kappa \mu \chi (T)}
\end{equation}%
Since $\chi (T)$ strongly depends on the temperature both the maximum
value and the shape of $P(w)$ strongly depend on the temperature. A typical
temperature dependence of the spin polarization described by Eq.~(\ref%
{P-temp}) is shown in Fig.~\ref{P-T}. If, as previously, we attempt to
interpret Eq.~(\ref{P-temp}) using a simple exponential dependence $P\propto
\exp (-w/L_{eff}(T))$ we will get a spurious temperature dependence of the
apparent spin-diffusion length $L_{eff}(T)$ (see Fig.~\ref{L-T}.), as was
inferred by Geresdi~\emph{et al.}~\cite{Geresdi}, demonstrating that
neglecting the backflow effect could lead to erroneous results.

We use several approximations in our description of the experimental
geometry, such as adopting a one dimensional model for what is a 3D problem 
and using boundary conditions at the N/S interface that assume only Andreev
reflection below the gap, hence neglecting processes above the gap 
. While these approximations may introduce some systematic errors, they are
unlikely to significantly affect the rate of spin polarization decay, which
determines the values of spin diffusion length. We also note that within the
same approximations, it is possible to obtain a complete set of data needed
for the determination of spin diffusion length from a single sample by
sequentially positioning the tip for PCAR\ measurements along the side of
the normal electrode, as shown in Fig.\ref{schematics}.

In summary, the backflow effect on spin diffusion and spin accumulation is
formulated as a consequence of preferential majority scattering near normal
metal - superconducting interface. It is found that spin current probed by
Andreev Reflection measurements gradually decays, as we increase the
thickness of the normal layer, revealing the scale of spin diffusion in the
normal metal. The measured spin diffusion length in gold of approximately
285 nm, more than two time larger than that one would have obtained using a
simple exponential fit. While our experimental results are described
specifically for a normal metal - superconducting interface, we emphasize
the role of boundary conditions, noting that qualitatively similar effects
would take place for normal metal - ferromagnetic interface as well, and
thus are relevant for other spin diffusion length measurement techniques.

The authors thank E.I. Rashba, P. Crowell, A.A. Golubov, and I.I. Mazin for
very helpful discussions and useful suggestions. This work was supported by
DARPA SpinS through ONR
Grant No. N00014-02-1-0886 and NSF Career Grant No. 0239058, at WSU (B.N.). 


\begin{thebibliography}{99}
\bibitem{lowbarrier} We assume a low barrier/high transparency interface.
\end{thebibliography}
\end{document}